\documentclass[aps,superscriptaddress,floats,showpacs,floatfix,onecolumn]{revtex4}
\usepackage{bm}
\usepackage{graphicx}
\usepackage{subfigure}
\usepackage{amssymb}
\usepackage{amsmath}
\usepackage{color}
\usepackage[a4paper,left=2.cm, right=2.cm, top=2cm,bottom=2cm]{geometry}
\begin{document}
\newcommand{\joerg}[1]{\textcolor{red}{#1}}
\newcommand{\janet}[1]{\textcolor{blue}{#1}}

\title{  Transitions and Multi-Scaling in  Rayliegh-Benard Convection.  Small-Scale Universality. }
\author{Victor Yakhot}
\affiliation{Department of Mechanical Engineering, Boston University, Boston, MA 02215, USA}
\affiliation{EXA Corporation, 55 Network dr., Burlington, MA, USA 01803, }

\date{\today}

\begin{abstract}
\noindent  
 Asymptotically large Reynolds number  hydrodynamic turbulence is characterized by multi-scaling of moments of velocity increments and spatial derivatives. With decreasing Reynolds number toward  $R_{\lambda}=R^{tr}_{\lambda}\approx 9.0$,  the anomalous scaling disappears in favor of  the "normal" one  and  close-to-Gaussian probability densities  [Yakhot \& Donzis, {\bf 119}, 044501 (2017)].  The nature of this transition and its universality are subjects of this work. 
 Here we  consider Benard convection ( Prandtl number $Pr=1$)  between infinite horizontal plates.      It is shown that in this system  the "competition"  between Bolgiano and Kolmogorov processes,   results in  small-scale velocity  fluctuations   
    driven by   effective "large-scale"  Gaussian random temperature field.  Therefore,  the  intermittent  dynamics of velocity derivatives  are similar  or even identical    to  that in homogeneous  and isotropic turbulence generated by  the large-scale random forcing.  It is shown that  low-Rayleigh number instabilities  make the problem much more  involved and may  lead to transition from Gaussian to exponential PDF of the temperature field.  The developed  {\it mean-field theory}  yielded   dimensional heat flux  $Nu\propto Ra^{\beta}$ with $\beta\approx  15/56\approx 0.27$, close to the outcome of Chicago experiment. These results   point  to an   unusual small-scale universality of turbulent flows.  It is also shown that at $R_{\lambda}\leq 9.0$, a  flow "remembers" its laminar background and, therefore, cannot be universal.

 
\end{abstract}
\keywords{}
\maketitle

\section{Introduction}

 \noindent Transition to turbulence is the series of processes by which a flow passes from regular or laminar to irregular or turbulent as the control parameter, usually the Reynolds number Re, is increased.   Formulated this way,  transition
 can be perceived  as emergence of chaotic solutions  out of  deterministic equations  of motion as   a process of formation, interactions  and instabilities of coherent structures.   Thus, it is a subject  of  theory of chaos, pioneered  by Lorenz  in 1963 who studied instabilities  in Benard convection cell Ref.[1].  In this paper we are interested in another kind of transition - the one between two different states of developed turbulence,  formulated  in [2]-[4].  \\

\noindent  Transition from laminar   to turbulent flow  was  discovered and analyzed by Osborn Reynolds in 1883, who   reported  emergence  of "sinuous" motions  in  steady  water flow  in a  strait pipe.  Introducing mean velocity 
 $U\propto \int_{0}^{L} rU(r)dr$,   Reynolds quantified the phenomenon in terms 
 of  dimensionless parameter $Re=UL/\nu$,  later called Reynolds number, 
 so that at $Re\leq Re_{tr}$ the flow was laminar, with a steady parabolic velocity profile $U(r)$.  At $Re\geq Re_{tr}$, he noticed appearance of  irregular or random fluctuations  ${\bf v}({\bf x},t)$ on a laminar background.   With increase of $Re>Re_{tr}$,  the amplitude and degree of randomness increased which made analysis  of the flow very hard. Interestingly, Reynolds was the first to suggest description of this  flow using statistical methods of theoretical physics. The transitional Reynolds number $Re_{tr}\approx 2000-14000$,  reported in 1883,   happened to   depend on geometry and quality of the pipe. It has been demonstrated later that in very smooth pipes the flow stayed laminar  at least up to $Re_{tr}\approx 10^{5}$ while,  depending on degree of wall-roughness,   stochastic components in  velocity field  have been   observed at $Re_{tr}\leq 1000$. 
 \\

One can envision  an  infinite  flow  generated by a random force at the scale $L$.  If the energy input is sufficiently small,  non-linear effects are weak  and the system can be described  as  a large  set  of almost independent realizations. Therefore,  according to CLT,  often,  it obeys close-to-Gaussian statistics, derived below from dynamic equations. With increase of  pumping power, non-linear interactions  between the modes  grow leading to deviations from Gaussianity which are especially violent  at the tails of the probability density functions (PDF) corresponding to the  large - amplitude fluctuations. 
  In this paper,  we are  interested in transitions  between  weakly and strongly coupled 
 states of a random flow,   characterized by different PDF's.

The magnitude of  $Re_{tr}$ can be obtained from solutions  to  the Navier-Stokes equations  with subsequent investigation of their 
stability.  This, in general, very complicated procedure ( for the most recent  review see Ref.[1)] is   dramatically simplified  in the {\bf inverse}  program, introduced below.
Since  at the scales $r<<L$,  turbulence is  isotropic and homogeneous,  the 
 "eddies"  on the scale $l \approx r$    move with effective velocity Refs.[3]-[4].
 \begin{eqnarray}
 v_{r,n}=[S_{n}(r)]^{\frac{1}{n}}
 \end{eqnarray}
 
 \noindent  Here, the structure functions $S_{n}(r)$ are defined as:

\begin{eqnarray}
S_{n}(r)=\overline{(v_{x}({\bf x})-v_{x}({\bf x}+r{\bf i}))^{n}}\equiv \overline{(\delta_{r}v)^{n}}\equiv\overline{{v_{r}^{n}}} \propto  (\frac{r}{L})^{\zeta_{n}}
\end{eqnarray}

\noindent   so that  $v_{rms}=v_{r,2}$. Thus,  since on the scale $r$ the effective viscosity Ref.[5]:

 $$\nu_{r}\approx v_{r}r \approx \sqrt{\overline{v^{2}_{r}}}r$$
 
\noindent  the  local  Reynolds number  defied  is
$ Re_{n,r}=v_{r,n}r/\nu$ is a strongly fluctuating flow characteristic which, in some realizations,   may be larger than the transitional one. This may lead to appearance of turbulence patches in sub-transitional flows. On the integral scale 
$\nu_{T}\approx 0.0845 {\overline{v^{2}}}^{2}/{\cal E}$
\noindent and the effective dynamic  or "dressed" Reynolds number based on Taylor scale is (see Table I) : 

$$R_{\lambda}\approx \sqrt{\frac{5}{3{\cal E}\nu_{T}}}v_{rms}^{2}\approx 9.0$$

\noindent  The result,  $R_{\lambda}\approx 10-12$,  has been experimentally observed in many engineering setups  like flows past  bluff bodies like spheres and cars   and in numerical simulations of decaying turbulence Ref.[10]. 
Also,  in recent numerical experiments on  isotropic and homogeneous turbulence (HIT),  
driven  by {\bf different}  random forces,  the transition from  the low-Reynolds number Gaussian  to strongly   anomalous turbulence  was found at $R_{\lambda}^{tr}\approx 9.0$, very close $Re_{tr}\approx 8.91$ resulting from the  RG -calculations of Refs. [ 6]-[10]. \\

\section{ Multitude of ``Reynolds numbers''  in a fully developed turbulent flow. Tails and rare events.}

 \noindent   
Let us identify velocity $v_{r,n}$,   defined in (1)-(2) ,   with  velocity   of local $n^{th}$-order   realization  on a scale $r$. Then, in the limit $r/L\ll 1$,  the ratio

$$\frac{v_{r,n}(K41)}{v_{r,n}(I)}=(\frac{r}{L})^{\kappa(n)}\rightarrow 0$$

\noindent where $\kappa(n)=\frac{3}{4}-\frac{\zeta_{n}}{n}$. Due to intermittency or anomalous scaling,  $\zeta_{n}<3n/4$ and, as $n\rightarrow \infty$,  $  \frac{v_{r,n}(K41)}{v_{r,n}(I)}\rightarrow 0$. Here, $K41$  and $I$ mark the outcomes  of  Kolmogorov and intermittency theories,  respectively.
This means that in high Reynolds number flows,  at the tails, PDFs of small-scale velocity increments  are much "fatter"   than the typical Gaussian PDF,  characteristic of K41.   Thus, in these  not -so-rare flow realizations,  corresponding to  extreme small-scale fluctuations,  the {\bf local} Reynolds number may be substantially larger than the one based on mild fluctuations with typical velocity $v_{rms}$. In principle, locally, the Reynolds number may be larger than that of transition leading to formation of  turbulent patches and dissipative structures. \\

\noindent The moments of  derivatives,    including   those of dissipation rate,  ${\cal E}=\nu(\frac{\partial v_{i}}{\partial x_{j}})^{2}$,   
 which are a plausible descriptor of small-scale dynamics,  are defined as:

$$M_{2n}=\frac{\overline{(\partial_{x}v_{x})^{2n}}}{{\overline{(\partial_{x}v_{x})^{2}}^{n}}}\propto Re^{\rho_{2n}}$$

\noindent where the large-scale Reynolds number $Re$ is defined in Table 1.  In the vicinity of transition point we define $v_{rms}\equiv v_{0}$ and the integral and dissipation scales $\eta\approx \lambda$.  Multiplying  and dividing $M_{2n}$ by $\nu^{n}$ gives: 

$$\overline{(\partial_{x}v_{x})^{2n}}=(\frac{v_{0}}{L})^{2n}Re^{2n}={\overline{\cal E}^{n}}\nu^{n}\propto Re^{d_{n}-n}(\frac{v_{0}}{L})^{2n}$$

\noindent leading  to an important relation [3-[4]]:

\begin{eqnarray}
\rho_{2n}=d_{n}+n 
\end{eqnarray}

\noindent Remarkably, the relatively sharp boundary separating    "anomalous"  and "normal"   scalings  of velocity derivatives  and increments,   is very close to $R_{\lambda}\approx 9.-10$. (Ref.[2]-[4]).  It has been shown recently that  in  infinite fluids,  stirred by different large-scale random forces,  
the exponents $d_{n}$ of  the moments of dissipation rate:

$$e_{n}
=\frac{\overline{{\cal E}^{n}}}{\overline{\cal E}^{n}}\propto Re^{d_{n}}$$

\noindent   are forcing-independent in the range $R_{\lambda}\geq 9.0-10.0$.    
In this paper we investigate transition between "normal" and anomalous flow regimes  in Benard convection between two infinite plates heated from below.

It became clear that the so-called Kolmogorov's scaling $\zeta_{n}=n/3$ and $\rho_{2n}=n$  is not valid for  $n\neq 3$ and the moments of orders $m$ and $n$ with $m\neq n$ 
are given by some "strange" numbers not  
related to each other by dimensional considerations.  This feature of strong turbulence, called "anomalous scaling", is the signature of strong interactions between different modes  in non-linear systems.
For many years theoretical evaluation of anomalous 
 exponents $\zeta_{n}$ and $\rho_{n}$ was considered one of the main goals   of the proverbial  "turbulence problem". 
 It was shown both theoretically and numerically  in Refs.[3]-[4] that possible reason for this difficulty is hidden in the fact that each moment $S_{n}(r)$  and  $M_{n}$ should be characterized by its "own" $n$-dependent Reynolds number $\hat{R}e_{n}$ based on characteristic velocity $ \hat{v}(n,n)$, defined in Table 1, and  that a widely used parameter $v_{rms}=\hat{v}(2,2)$ is simply one of an infinite number of characteristic velocities describing turbulent flow. 
 The multitude of  dynamically relevant Reynolds numbers,  necessary for description of turbulence,  is  defined in Table 1.

\begin{table}[h]
\begin{tabular}{l|l}
\hline
Reynolds number & Description \\ \hline \hline
$v_{rms}=\sqrt{\overline{v^{2}}}$ & root-mean-square velocity\\
$\hat{v}(m,n)=\overline{|v|^{m}}^{\frac{1}{n}}$ &moment of order $m/n$;  $v_{rms}=\hat{v}(2,2)\equiv \hat{v}_{2}$\\
$Re =v_{rms} L/\nu$ & large-scale Reynolds number\\
$\hat{Re}_{n}=\hat{v}(n,n)L/\nu$ & Reynolds number  of the   $n^{th}$ moment \\
 $R_\lambda = v_{rms} \lambda/\nu$ & Taylor Reynolds number;
                          $\lambda=15\nu u_{rms}^2/{\cal E}$  \\
$Re_{n}^{tr}$ & transition point 
                     for moments of order $n$ \\
$\hat{R}e_n=\hat{v}(n,n)L/\nu$ &  
probes regions with different amplitudes of velocity gradients\\
$\hat{R}_{\lambda,n}=(5L^4/3{\cal E} \nu)^{1/2}\hat{v}(2n,n)$ &  
 order-dependent Taylor-scale Reynolds number\\
\\
\hline
\end{tabular}
\label{tab:res}
\end{table}

\begin{figure*} 
\includegraphics[height=6cm]{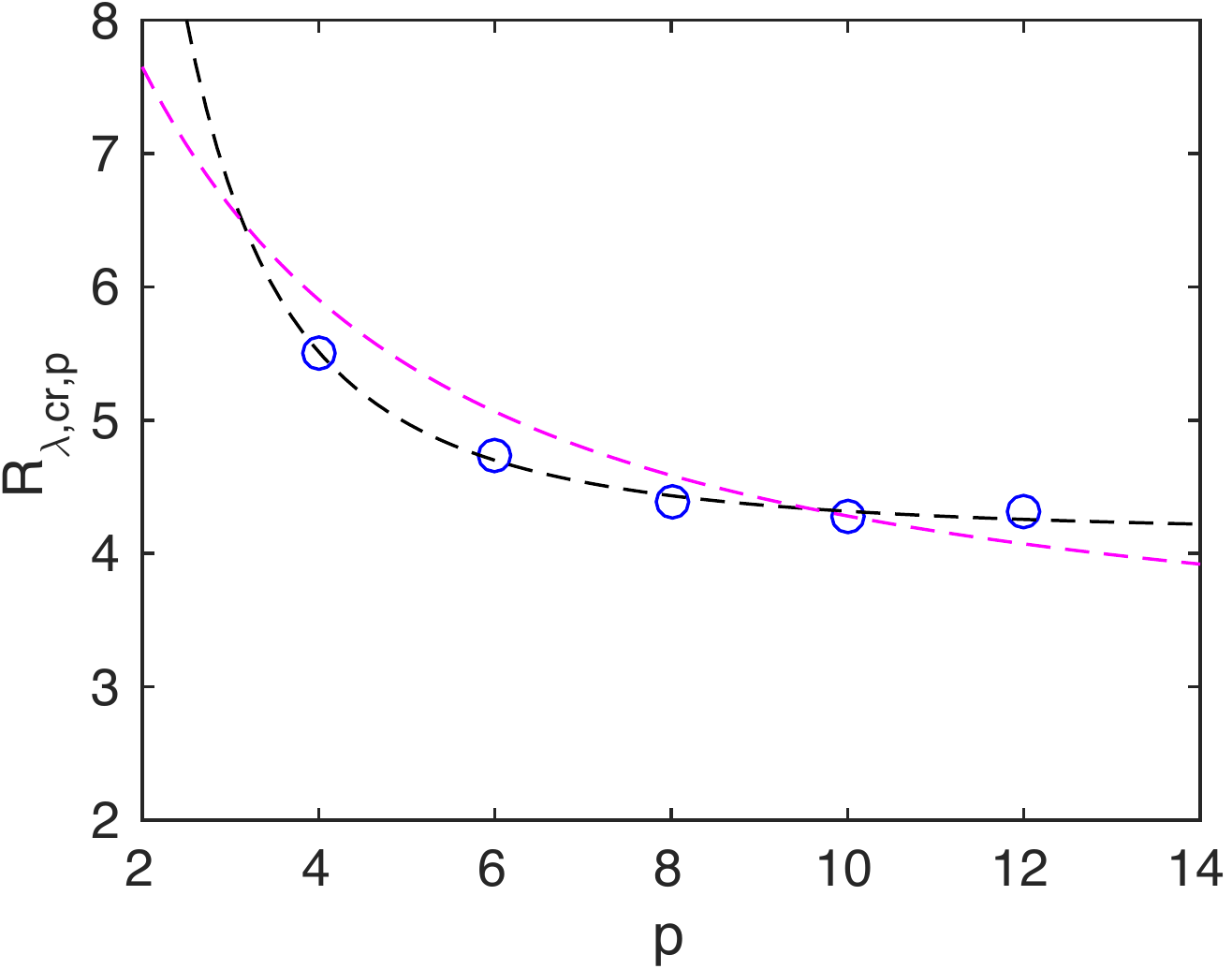}
\caption{Moment-Order-Dependent Transitional Reynolds Numbers. Circles: predicted and numerically tested in Ref.[3]-[4]. Dotted line: numerical simulations  of the flow driven by the large-scale random force of Refs. [3]-[4].   Reynolds number of transition decreases with increase of the moment order of $M_{n}$.   This explains appearance of turbulence patches/puffs in  ``sub -transitional'' flows. }
 \end{figure*}

\noindent  We would like to stress an important point: $v_{rms}$  characterizes   typical or  relatively mild velocity fluctuations. In general, to be able to predict   rare,  extreme,  events  we   introduce   $\hat{v}_{2n}=L^{2}\overline{(\partial_{x}v_{x})^{2n}}^{\frac{1}{n}}\propto A_{2n}^{\frac{1}{n}}Re^{\frac{\rho_{2n}}{n}}$ and $\hat{R}_{\lambda, n}^{tr}=\sqrt{\frac{5}{{3\cal E}\nu}}\hat{v}_{2n}\approx 8.91$ derived in Refs.[6]-[9].  To calculate  large- scale transitional Reynolds number  we introduce velocity scale $v_{0}=v_{rms}$  so that $Re=v_{0}L/\nu$ and :

$$\hat{R}_{\lambda,n}^{tr}=\sqrt{\frac{5}{3{\cal E}\nu}}\hat{v}_{2n}=A_{2n}^{\frac{1}{n}}(Re^{tr})^{\frac{\rho_{2n}}{n}+\frac{1}{2}}\approx 8.91$$

\noindent It follows from this relation that transition to strong turbulence in different realizations or different-order-moments occurs at a constant $R^{tr}_{\lambda,n}=8.91$  but at {\bf different}  $Re^{tr}=v_{rms}L/\nu$ based on the r.m.s. velocity coming from the second-order moment. This result, theoretically evaluated in [7]-[9],  is consistent with the empirical   ${\cal K}-{\cal E}$ model 
giving the large-scale ``dressed'' viscosity $\nu_{T}=0.0845{\cal K}^{2}/{\cal E}$,   used in  engineering simulations during last fifty years [6]. Indeed: with ${\cal K}=v_{rms}^{2}/2$

$$R^{tr}_{\lambda}\equiv R^{tr}_{\lambda,2}=\sqrt{\frac{5}{3{\cal E}\nu_{T}}}2{\cal K}\approx 8.88$$

\noindent  and 

$$Re_{n}^{tr}=[\frac{\hat{R}_{\lambda,n}^{tr}}{A_{2n}^{\frac{1}{n}}}]^{\frac{2\rho_{2n}}{2\rho_{2n}+n}}$$

The somewhat  ``unexpected'' but qualitatively reasonable  consequence of this result,  is  seen on Figs.2-3,  where the  onsets  of anomalous 
scaling for different moments $M_{n}$  are observed at very different $Re_{n}^{tr}$ but at a  {\bf single} $n$-independent  $\hat{R}_{\lambda,n}^{tr}\approx 9.0-10$.  For large enough $n$,  $A_{2n}^{\frac{1}{n}}$ is a weakly dependent function of $n$ which can be calculated from the $Re _{2}^{tr}\approx 9-10$. 
Thus, one can easily express    $Re_{n}^{tr}$ in terms of $\hat{R}_{\lambda,n}^{tr}\approx 9.0-10$ [3] - [4] and close the equation (4) for $\rho_{2n}$.  The results are presented in Table II  and compared with the data on  Fig.1.

\section{ Two examples: flows driven by  large-scale external random forces.} 
 
\noindent In Landau's theory of "laminar-to-turbulent transition", the Reynolds number is defined on a ``typical'' characteristic  velocity $V$ and length-scale $L$ depending   on flow geometry, dimensionality, 
physical mechanisms responsible for  instability and other factors characterizing large-scale ordered (laminar) flow.  Therefore, in this  approach $Re_{tr}$  varies in an extremely  wide range of parameter variation. 
To study  dynamics of velocity fluctuations  it is useful to define the   Reynolds number   $Re=v_{rms}L/\nu=\sqrt{\overline{v^{2}}}L/\nu$ based entirely on fluctuating velocity  ${\bf v}$ for which $\overline{\bf v}=0$.  
To avoid difficulties  related to instabilities of a laminar flow,   we studied  the dynamics governed by  the  Navier-Stokes equations  in an infinite fluid stirred by a Gaussian random forcing  acting on a finite scale $r\approx L$  Refs. [3]-[4]:

\begin{equation}\partial_{t}{\bf v}+{\bf v\cdot\nabla v}=-\nabla p +\nu\nabla^{2}{\bf v} + {\bf f} \end{equation}

\noindent   $\nabla\cdot {\bf v}=0$.  Here the density is taken $\rho=1$ without loss of generality.  A random Gaussian noise ${\bf f} $ is  defined by  correlation function:

 \begin{equation}\overline{f_{i}({\bf  k},\omega)f_{j}({\bf k'},\omega')}= (2\pi)^{d+1}D_{0}(k)P_{ij}({\bf k})\delta({\hat{k}+\hat{k}'})\end{equation}

\noindent  where the four-vector $\hat{k}=({\bf k},\omega )$ and projection operator is: $P_{ij}({\bf k})=\delta_{ij}-\frac{k_{i}k_{j}}{k^{2}}$.    It is clear from (4)-(5) that in the limit $D_{0}\rightarrow 0$ the nonlinearity is small and ${\bf v}(\hat{k})\approx G^{0}{\bf f }=O(\sqrt{D_{0}})$,  where the ``bare'' Green function is  $G^{0}=1/(-i\omega +\nu k^{2})$. In this limit the velocity field is Gaussian with  the derivative moments $M_{2n}=\overline{(\partial_{x}v_{x})^{2n}}/\overline{(\partial_{x}v_{x})^{2}}^{n}\approx (2n-1)!!$. \\

\noindent The second example is the one of the Navier-Stokes equations  driven by   a different  forcing mechanism 
in the rhs of (2) defined as: 

\begin{equation}
{\bf f}({\bf k},t)=\overline{{\cal E}}\frac{{\bf u}({\bf k},t)}{\sum_{k=1,2}|{\bf u}(\hat{k})|^{2}}\delta_{\bf k,k'}
\end{equation} 

\noindent where $\hat{k}=({\bf k},t)$  is a self-consistent solution to the NS equation   not equal to zero only at  $k_{i}=1;2;$. 
The advantage of this force is the fixed rate (power)  of "turbulence production" mechanism $\overline{{\bf f }\cdot {\bf u}}=1.0$.  The results of numerical simulations are presented on the  left and right panels of Fig.2. We see that  in both cases transition  
from Gaussian to anomalous scalings occurs  at  theoretically predicted $R_{\lambda}\approx 9.0$. 
In the interval $R_{\lambda,n}>R^{tr}_{\lambda,n}$ numerical simulations  gave  $M_{2n}\propto R_{\lambda}^{\rho_{n}}$ with anomalous exponents $\rho_{n}$ shown in Table.I. (also see Ref.[2]).

\noindent  {\bf 
Both examples, considered above,   dealt with an infinite fluid  stirred at a finite scale $L=O(1)$ . This means that if  linear dimension of a fluid is ${\cal L}\rightarrow  \infty$, then the flow is generated by $N = {\cal L}^{3}/L^{3}\rightarrow \infty$ random, uncorrelated, stirrers, each one defining  a statistical realization. 
  Therefore, one can  describe a  flow either in terms of fluctuations  of local parameters or, equivalently, by statistical ensemble with corresponding  probability densities (PDFs). This will be demonstrated  in detail below.}\\

\begin{figure*} 
\includegraphics[height=8cm,width=7cm]{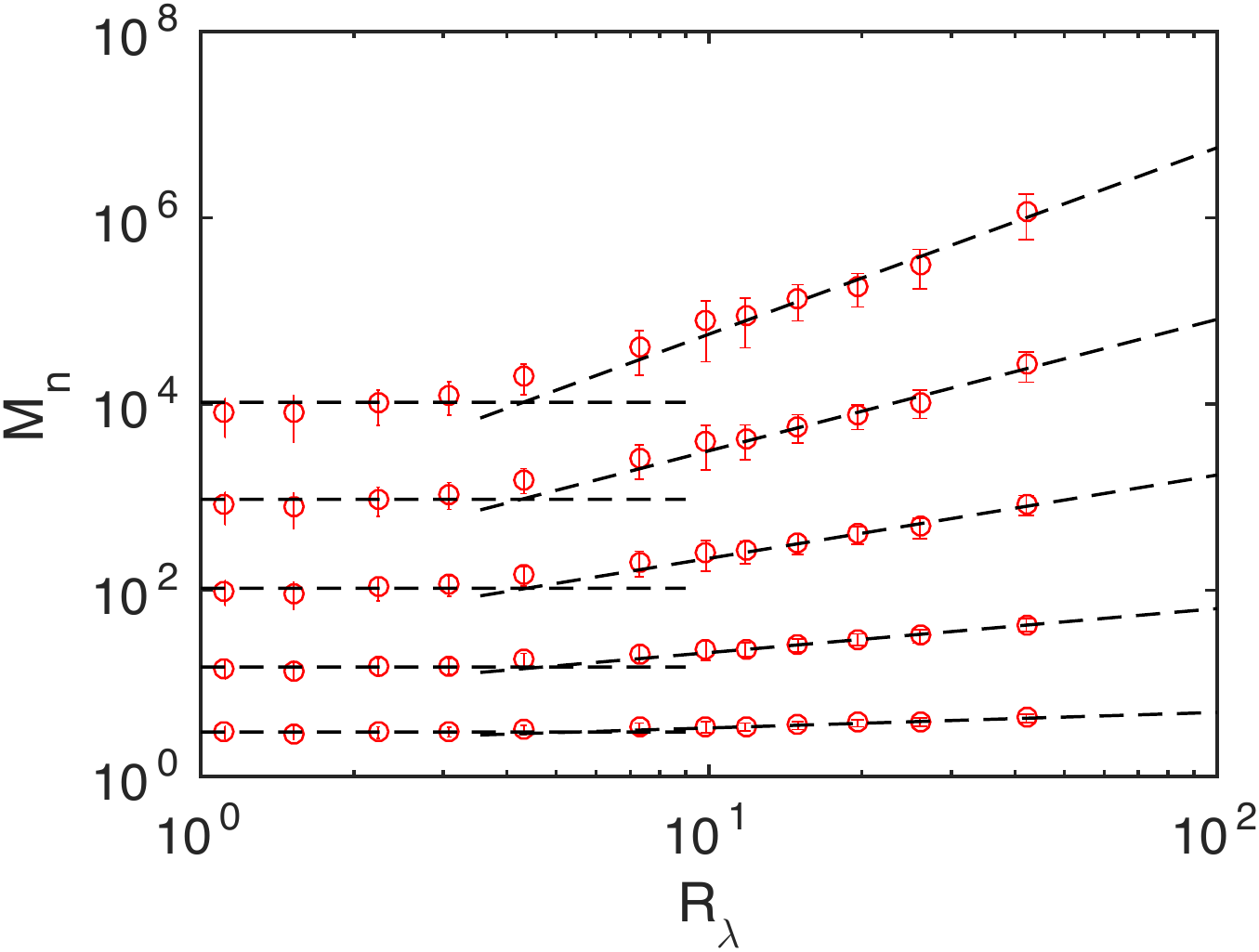}
\includegraphics[height = 8cm,width=8cm]{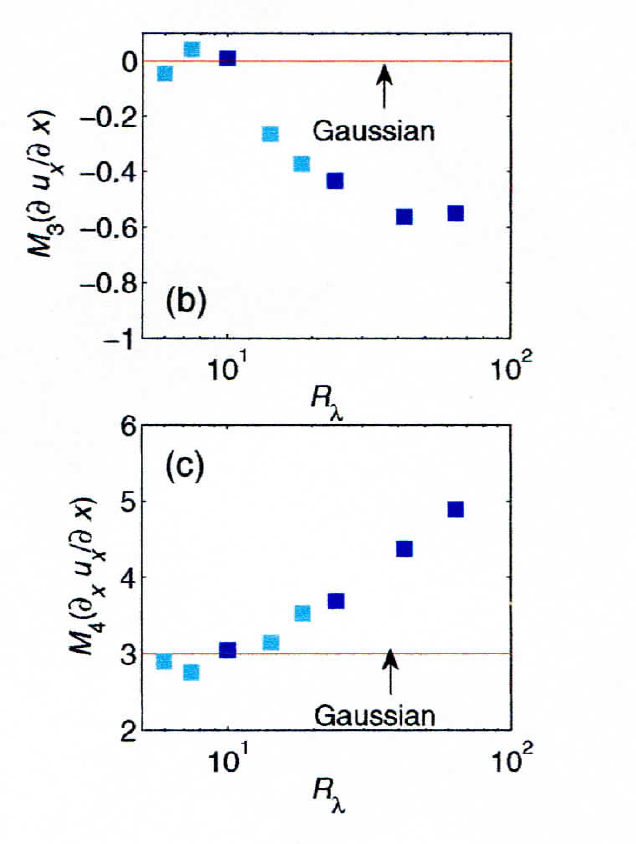}
\caption{ Left   panel:  Normalized moments of velocity gradients $M_{2n}$ from direct numerical  solutions to the Navier-Stokes equations (4)-(5).  At $R_{\lambda}<R^{tr}_{\lambda,n}$ the moments $M_{2n}=(2n-1)!!$, i.e.  obey Gaussian statistics.   In the range $R_{\lambda,2n}>9.0$, the moments $M_{2n}\propto R_{\lambda}^{\rho_{n}}$, where $\rho_{n}$ are anomalous scaling exponents  from Ref.~[2]-[4]. Right panel: flow driven by fhe forcing (6) studied in Ref.[2].  The transition from Gaussian to "anomalous"  turbulence at  the same $R_{\lambda}\approx 9$ is clearly seen. Sub-gaussian fluctuations in the range  $R_{\lambda}<9$  correspond to the more complex low-$R_{\lambda}$ dynamics reflecting complexity of the forcing Ref.[2].}
\end{figure*}

 \begin{table}
\begin{ruledtabular}
\begin{tabular}{ccccccccccc}
\hline
$ \rho_{n} $ & $ GAU$& $DNS$\\
\hline
$\rho_{1} $&  $0.46$& $0.455$\\
$ \rho_{3}$ & $1.58$ & $1.478$\\
$ \rho_{4}$ & $2.19$ & $2.05$\\
 $ \rho_{5}$&  $2.82 $& $2.66\pm 0.14$\\
 $ \rho_{7} $ & $4.13$ & $3.99\pm 0.65$\\
 \end{tabular}
 \end{ruledtabular}
\caption{Comparison of exponents $\rho_{2n}=d_{n}+n$  with the outcome of numerical  simulations (DNS)  and  a theory  [3].       }
\end{table}
 \begin{table}
\begin{ruledtabular}
\begin{tabular}{ccccccccccc}
\hline
$n$ & $1 $ &  $ 2 $ & $3$ & $4$ \\
$R_{\lambda,2n}^{tr}$  & $8.91$ & $5.5$ &  $4.8$ & $4.5$  \\
$Re_{2n}^{tr}$ & $126$ & $45$ & $35 $ & $ 30 $\\
 \end{tabular}
 \end{ruledtabular}
\caption{ Transitional Reynolds numbers based on Taylor scale $R^{tr}_{\lambda,2n}=\sqrt{\frac{5}{3{\cal E}\nu}}v_{rms}^{2}$  of the moments $M_{2n}$. 
With $\hat{v}_{2}=\overline{v^{2n}}^{\frac{1}{n}}$, the  modified Reynolds number $\hat{R}_{\lambda,2n}^{tr}=8.91$ is  independent on $n$.}
\end{table}


 \section{  Thermal Convection. }
 
 \noindent The simplified problems of Refs. [2]-[4], described in a previous Section,  dealt with transition between a 
  Gaussian  state of a  fluid  and the  non-linearity-dominated strong, anomalous,  turbulence.  In each case,   
  flow was driven by an externally prescribed  random force. 
   In real-life -flows,  various   randomness - generating mechanisms often act   simultaneously: for example  in wall flows 
 turbulence is  produced  by   instabilities  of both  quasi-laminar flow patterns in the bulk and those of  viscous wall layers,  generating   powerful bursts reaching  into  the bulk of  a flow. Thus, in this case, the mechanism of transition to anomalous scaling may be much more involved.
 Below, based on the theory developed in Refs.[12]-[13],  we address this problem.   The time-dependent number of theoretical, experimental and numerical publications  on stability of a fluid heated from below, including Benard convection,   is enormous and below we restrict ourselves by quoting only a few necessary sources the  derivations are  based on. \\

\subsection{Phenomenology.  Reynolds and Rayleigh numbers.}

\noindent  We are interested in  a  small-scale behavior of a flow between two infinite,  thermally insulated,  plates separated  by the gap $H$. The bottom  plate  at $z=0$  is heated by an electric current $I$.  Sufficiently far from    thermal boundary layers,   the heat flux averaged over horizontal planes $J_{T}(z)=J(T)=const$ and we keep  top and bottom plates under  constant temperature difference $T_{top}-T_{bot}\equiv \Delta=const$.  \\

\noindent  Transition to asymptotic, high Rayleigh number  limit, involving formation and  transformations of various large-scale coherent structures  is an extremely complex process Ref.[1], [14]-[16],  [17]-[18].   Detailed study of this   fascinating  chain of events  is beyond the scope of this paper.  Here we restrict ourselves  to investigation  of a few basic steps. First,  in  the heat conduction range, $Ra= \frac{\alpha g H^{3} \Delta}{\kappa\nu}  \leq 1708$,  the temperature gradient results in  heat  flux $J_{T}=\kappa \Delta/H$. In this regime the velocity field ${\bf v} =0$.  At $Ra\geq 1708$,  the first rolls,  contributing  to  the heat flux,  appear on a scale $r\approx H$.  In the interval  $10^{5}\leq Ra\leq 10^{7}$,  
 according to Busse Ref.[ 15]  and  Krishnamurti Ref.[14], the flow consists of rolls and hexagonal cells   coexisting  with imbedded small-scale velocity fluctuations, they call "convection  elements".  These fluctuations  exist on the  length-scales $r\leq H$. In the interval   $Ra> 10^{8}$, the instability of  boundary layers  leads to formation of powerful plumes emitted into  the bulk and eventually dominating heat transfer process  Ref.[18].  At larger Rayleigh numbers,  the fully turbulent  flow is dominated by non-linear terms in the equations of motion and in this limit we may expect some kind of universality.

\noindent We consider  the coupled three-dimensional equations of motion for velocity and  temperature fluctuations $v_{i}$  and $T$, respectively:  
\begin{align}
\label{nseq}
\frac{\partial v_{i}}{\partial t}+v_{j} \frac{\partial v_{i}}{\partial x_{j}}
&=-\frac{\partial p}{\partial x_{i}}+\nu \frac{\partial^2  v_{i}}{\partial x_{i}{^2}}+  \alpha gT \delta_{i3}\,,\\
\frac{\partial T}{\partial  t}+v_{j} \frac{\partial T}{\partial x_{j}},
&=\kappa \frac{\partial^2 T}{\partial x_{j}^2}+\kappa\frac{\partial^{2} \Theta}{\partial x_{j}^{2}}-v_{3}\frac{\partial \Theta}{\partial x_{3}}\,,
\label{pseq}
\end{align}

\noindent Here the horizontally averaged  temperature  $\Theta=\Theta(z)$  and $\Theta(0)-\Theta(H)=\Delta$. Neglecting  compressibility,  we  set $\partial_{j}v_{j}=0$. 
Outside thermal boundary  $0\leq z\leq l_{B}\equiv l_{d}^{T}$, where both temperature and velocity fluctuations are statistically isotropic field and 
the  mean temperature $\Theta(z)\approx constant$,  the heat flux $J_{T}(z)\approx \kappa\frac{\Delta}{l_{B}} \approx const. $.
Here $l_{B}$ an approximate thickness of thermal boundary layer which is to be found from equations  (7)-(8).\\

\noindent Superficially,  the equation (7) looks like equation  (4)-(5),  considered in Ref.[3]-[4] , but with  the forcing ${\bf f}=\alpha g T{\bf k}$, where ${\bf k}(0,0,1)$.  This difference is profound:  while ${\bf f}$, defined in (5) is,  a large-scale Gaussian  process, the forcing in e.g..(7) is a solution to  dynamic equation (7)-(8). Therefore,  first, we would like to investigate temperature fluctuations described by (8).  We define:
$$
S_{n}(r)\approx \overline{(v(x+r)-v(x))^{n}}\equiv \overline{(\delta _{r}v)^{n} } ;       \hspace{2cm}  S^{T}_{3}(r)\approx \overline{\delta_{r}v (\delta_{r}T)^{2}}
$$

\noindent  One can easily derive the  balance relations:

\begin{eqnarray}
S^{T}_{3}(r)\approx -\frac{4}{3}Nr+\frac{2}{r^{2}}\int_{0}^{r}y^{2}\overline{\delta_{r}v_{3}\delta_{r}T}dy\frac{\partial \Theta}{\partial z}+2\kappa\frac{\partial S^{T}_{2}(r)}{\partial r}\\
S_{3}(r)\approx-\frac{4}{5}\epsilon r+\frac{2\alpha g}{r^{4}}\int_{0}^{r}y^{4}\overline{\delta_{r}v\delta_{r}T}dy+6\nu\frac{\partial S_{2}}{\partial r}
\end{eqnarray}

\noindent  In dimensionless variables: ${\hat v}=v/v_{rms}; \  {\hat{r}=r/H;  \hat{\cal E}=\frac{{\cal E}H}{v_{rms}^{3}}}; 
\hat{N}=\frac{NH^{2}}{\kappa \Delta^{2}}$  with 
$Ra=\frac{\alpha g H^{3} \Delta}{\nu^{2}}$, the relations (9)- (10) are:

\begin{eqnarray}
\hat{S}_{3}=-\frac{2}{15}\hat{\cal E}\hat{r} +\frac{12}{\hat{r}^{4}}\frac{Ra}{Re^{2}}\int_{0}^{r}\hat{y}^{4}\overline{\delta_{r}\hat{v}\delta_{r}\hat{T}}d\hat{y}+\frac{1}{Re}\frac{\partial \hat{S}_{2}}{\partial \hat{r}}\nonumber\\
\hat{S}^{T}_{3}=-\frac{4}{3}\hat{\cal N}\hat{r} +\frac{2}{\hat{r}^{2}}\int_{0}^{\hat{r}}\hat{y}^{2}\overline{\delta_{r}\hat{v}\delta_{r}\hat{T}}d\hat{y}\frac{\partial \hat{\Theta}}{\partial \hat{z}}+\frac{\partial \hat{S}_{2}}{\partial \hat{r}}\end{eqnarray}

\noindent It will be shown below that as  $Ra\rightarrow \infty$, the ratio  $Ra/Re^{2}\rightarrow \infty$ and $\hat{{\cal E}}=O(1)$ in accord with Kolmogorov relation for isotropic turbulence. 
\\   
 

 \subsection{Statistical ensemble.  Probability Density $P(X)$.  Low "Reynolds Number".}

\noindent   In  an infinite flow driven by  externally  prescribed large-scale random noise,  described by the equations (4)-(6),  one deals with a single transition between  a Gaussian    ($R_{\lambda}\leq R_{\lambda,tr}$)  and  anomalous    ($R_{\lambda}\gg  R_{\lambda, tr}$)   states.  It was shown that,  in the two examples discussed above, the transitional parameter  $R_{\lambda,tr}\approx 9.0$, 
independently of the nature of  forcing (See Fig.2).   This is not a universal statement: in wall flows,  in addition to processes happening in the bulk,   instabilities of  viscous and thermal boundary layers ($z\leq l_{B}$)  result in formation of powerful plumes and bursts reaching the bulk  and  supplying  a substantial contribution to   the fluxes of  kinetic energy  and heat. It is clear that  geometric details of a cell, like side walls, aspect ratios, curvature play an important part. It  will be shown below that each instability of this kind  is reflected in the shape of probability distributions functions of temperature and velocity fluctuations.  As  follows from  (7)-(8):  $\overline{{\cal E}}=-\alpha g\overline{v_{3}T}$
and  our goal is  evaluation of all  moments $e_{n}=\frac{\overline{{\cal E}^{n}}}{(\overline{(v_{3}T)}^{n}}$.  Introducing   dimensionless temperature:

$$ X=\frac{T'}{\Delta}=\frac{V_{pl}\sqrt{\cal E}}{\Delta\sqrt{\nu}\alpha g}=\sqrt{{\cal E}/\overline{{\cal E}}}\equiv\sqrt{e} $$

\noindent  we would like to evaluate the moments  $\overline{X^{2n}}\approx \overline{e^{n}}$.

 \noindent Below we use a somewhat modified  theory of a passive scalar proposed in  [11]  and applied to the problem of  Benard convection  in Ref.[12].  
 Multiplying (8) by $T^{2n-1}$  and,    since  $\overline{ \partial_{j}(v_{j} T^{2n})}=0$, we derive readily:

$$-(2n-1)\overline{T^{2n-2}(\nabla T)^{2}}=\overline{T^{2n-1}v_{3}}\frac{\partial \Theta}{\partial z}$$

\noindent With $X^{2}=T^{2}/\overline{T^{2}}$, $Y^{2}=(\nabla T)^{2}/\overline{(\nabla T)^{2}}$ and $W=v_{3}T/\overline{v_{3}T}$.  
These equations can be rewritten:

$$(2n-1)\overline{X^{2n-2}Y^{2}}=\overline{X^{2n-2}W}$$

\noindent and introducing conditional expectation values,  gives [10]-[12]:

$$(2n-1)\int X^{2n-2}r_{1}(X)P(X)dX=\int X^{2n-2}r_{3}(X)P(X)dX $$

\noindent where 

$$r_{1}(X)=\frac{\int Y^{2}(x)\delta(X(x)-X)dx}{\int \delta(X(x)-X)dx} \hspace{2cm} r_{3}(X)=\frac{X}{\overline{v_{3}X}} \frac{\int v_{3}(x)\delta(X(x)-X)dx}{\int \delta(X(x)-X)dx}$$

\noindent  and $r_{1}(X)$  and $r_{3}(X)$ are conditional expectation values of temperature dissipation and production  rates for fixed 
magnitude of dimensional temperature $X$. After simple manipulations, taking into account that at small $X$ production $r_{3}\propto \frac{X^{2}}{\overline{v_{3}X}}$,  one obtains a formally exact  representation of  
 probability density $P(X)$ [12]-[13]:

\begin{eqnarray}
P(X)=\frac{C}{r_{1}(X)}\exp\big[-\int_{0}^{X}\frac{r_{3}(u)du}{ur_{1}(u)}\big ] = \frac{C}{r_{1}(X)}\exp\big[-\int_{0}^{X}\frac{uv_{3}(u)du}{ur_{1}(u)}\big ]
\end{eqnarray}

\noindent Interested in the  low  $Re\leq Re_{tr}\approx 9.0$  limit,  we  evaluate  this expression in the limit $X\rightarrow 0$.  First, according to [11]-[12],  positive definite conditional dissipation rate 

 $$r_{1}(X)\approx \alpha +\beta X^{2}=\alpha(1+\frac{\beta}{\alpha}X^{2}) $$

\noindent  Since  $v_{3}(T)\approx -v_{3}(-T)$,  as $X\rightarrow 0$, we write $v_{3}(X)\approx \gamma  X$ and: 

\begin{eqnarray}
P(X)=\frac{C}{\alpha}e^{-\frac{\gamma}{2\alpha} X^{2}}
\end{eqnarray} 

\noindent and in the limit $X\rightarrow 0$,  the moments of dissipation rate at the center of the cell: $\overline{e^{n}}\approx (2n-1)!!$
similar to the moments of the dissipation rate  in isotropic and homogeneous turbulence generated by the large-scale forcing considered in Section III (See Refs.[2],[-4])\\

\begin{figure*}[h] 
 \includegraphics[height =5cm]{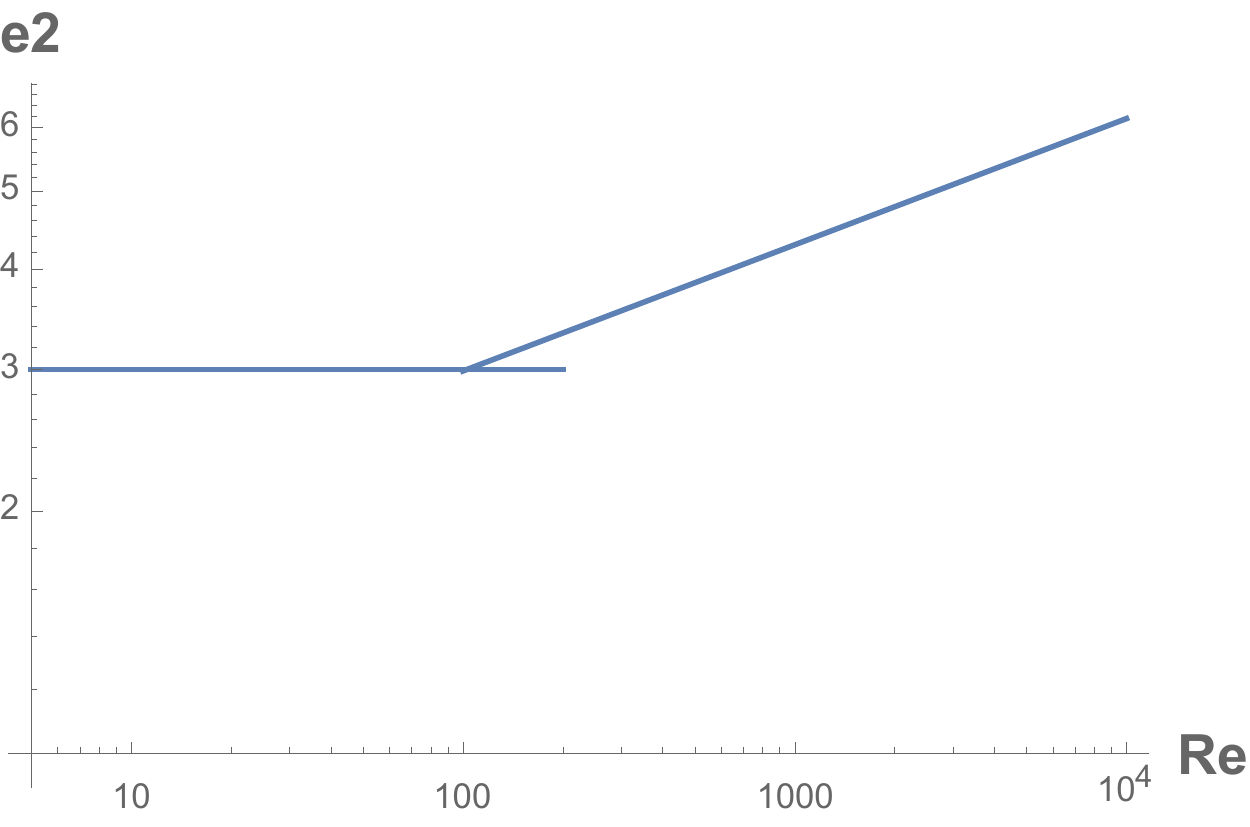}
\includegraphics[height=5cm]{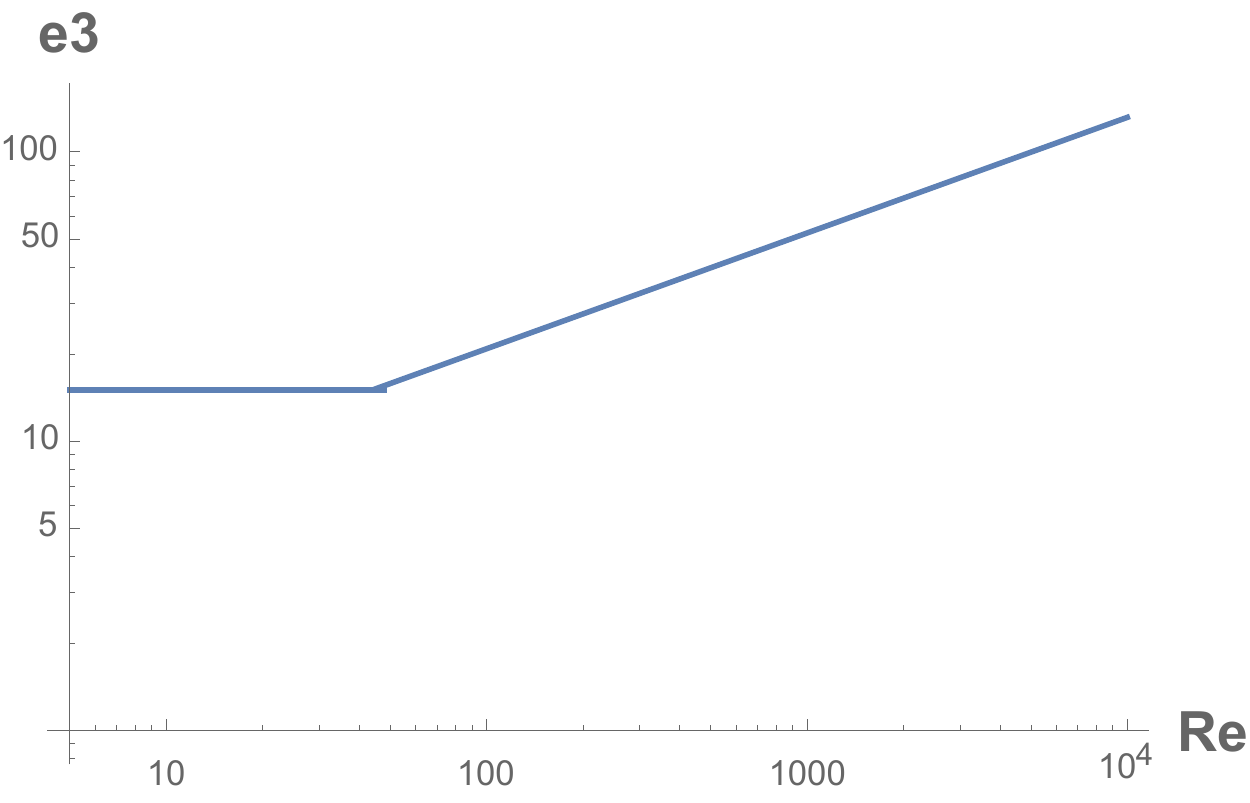}
\caption{ Normalized variance (left panel) and third-order moment of the dissipation rate fluctuations  (right panel) in the entire interval $Re\geq 10$. In the low  $Re\leq Re^{tr}_{n}$ range, the moments $e_{n}$  are equal to $e_{n}=(2n-1)!!$ in accord with (18).  The derived transitional $Re^{tr}_{n}$ are presented in Table  II.  In this case $Re_{pl}\approx 100$, so that $y=0$. The exponents  $d_{2}\approx 0.157$ and $d_{3}\approx 0.45$ are found from  the  matching conditions at  $(Re^{tr}_{n})^{d_{n}}=(2n-1)!!$.  The results are similar to those in HIT shown on Fig.2.  }
\end{figure*}

\subsection{Effect 
 of solid walls and flow   instabilities.}

\noindent In a finite cell, one has to deal with instabilities of boundary layers resulting in formation of plumes emitted into the bulk.  Also, formation of  "corner vortices",  fluctuations coming from the    side walls  and various  geometric details  may contribute to processes  of turbulence production and dissipation.
In a general case,  denoting  critical Reynolds numbers of various instabilities as $R_{pl}^{i}$,  we, following Ref.[12],  can write an expansion:

$$e\approx  X^{2}+\sum y_{i}X$$

\noindent where $y_{i}=(Re-Re^{i}_{pl})H(Re-Re^{i}_{pl})$  and $H(x)$ is Heaviside theta function. 
  One can see that, depending on the relative magnitudes of parameters $y_{i}$  and $X$, the heat transfer process is influenced by 
  different instabilities  contributing to the "bumps" in the $Nu(Ra)$ curve, observed in Chicago experiment  Ref.[18] .  Also, this result agrees with with the recent experimental data by Hong.[16]. Below, to simplify presentation,  we consider $i=1$ only and drop the subscript $i$. 
  Thus, the problem is reduced to evaluation of the moments:

\begin{eqnarray}
\overline{e^{n}}=\overline{(X^{2} +yX)^{n}} 
\end{eqnarray}

\noindent  Then,  since ${\bf v}_{3}\cdot  {\bf V}_{pl}>0$, we can write  $v_{3}=\gamma X +2yV_{pl}$, where $y=R_{\lambda}-R_{\lambda, pl}\geq 0$  and:

$$\frac{v_{3}(X)}{v_{rms}}\approx  \gamma X+2yV_{p}/v_{rms} \approx \gamma X+2\kappa y$$

\noindent where $V_{p}/v_{rms}=O(1)$.  We can see that when $y=0$, the resulting Gaussian flow is dominated by the weak "small-scale elements" [14]-[15]. 
Substituting all this into (12) gives:

$$P(X,y)=\frac{C(y)}{(1+\frac{\beta}{\alpha} X^{2})}\exp[\big[-\int_{0}^{X}\frac{\gamma u+2\kappa y}{\alpha(1+\frac{\beta}{\alpha} u^{2})}du \big ]$$ 

\noindent and the probability density of dimensionless temperature fluctuations at the central part of convection cell with $\alpha=\kappa=1$ is:
\begin{eqnarray}
P(X,y)=\frac{C(y)}{(1+ \frac{\beta}{\gamma}X^{2})^{1+\frac{\gamma}{2 \beta}}} exp(-2 y \arctan(\sqrt{ \beta}X)\equiv \nonumber\\
C(y)\Pi(X,y)
\end{eqnarray}

\noindent  with    $C(y)=1/ 2\int_{0}^{\infty}\Pi(X,y) dX $ 
and  $\beta\approx 1.4$ estimated in [12].  As $y\rightarrow 0$,  this expression gives Gaussian PDF with the half-width $\delta\approx \sqrt{\gamma/\beta}$.   
An interesting feature of this expression is the dependence of the  PDF on the Reynolds number $y\neq 0$, consequence of possible instabilities   of  boundary layers
in the range  $y>0$ and $R_{\lambda,2}\leq 9.0$.  In this range,  qualitative dependence  of PDF as a function of 
``Reynolds number'' $y$ is shown on Figs.4-5.

\subsection{Moments of dissipation rate. Low-Re regime.}

\noindent Based on the above derivation (also see Ref.[12]),   the conditional expectation value  of kinetic energy dissipation rate is approximated by the expression:

\begin{eqnarray}
\frac{\cal{E}}{\overline{\cal E}}\approx yX+X^{2}
\end{eqnarray}

\noindent  and thus, the normalized  moments of the dissipation rate are calculated readily

$$   e_{n}(y)= \frac{\int_{0}^{\infty} (yX+X^{2})^{2n})P(X,y)dX}{(\int_{0}^{\infty} (yX+X^{2})^{2})P(X,y)dX)^{n}}$$

\begin{figure}
\includegraphics[scale = 0.4]{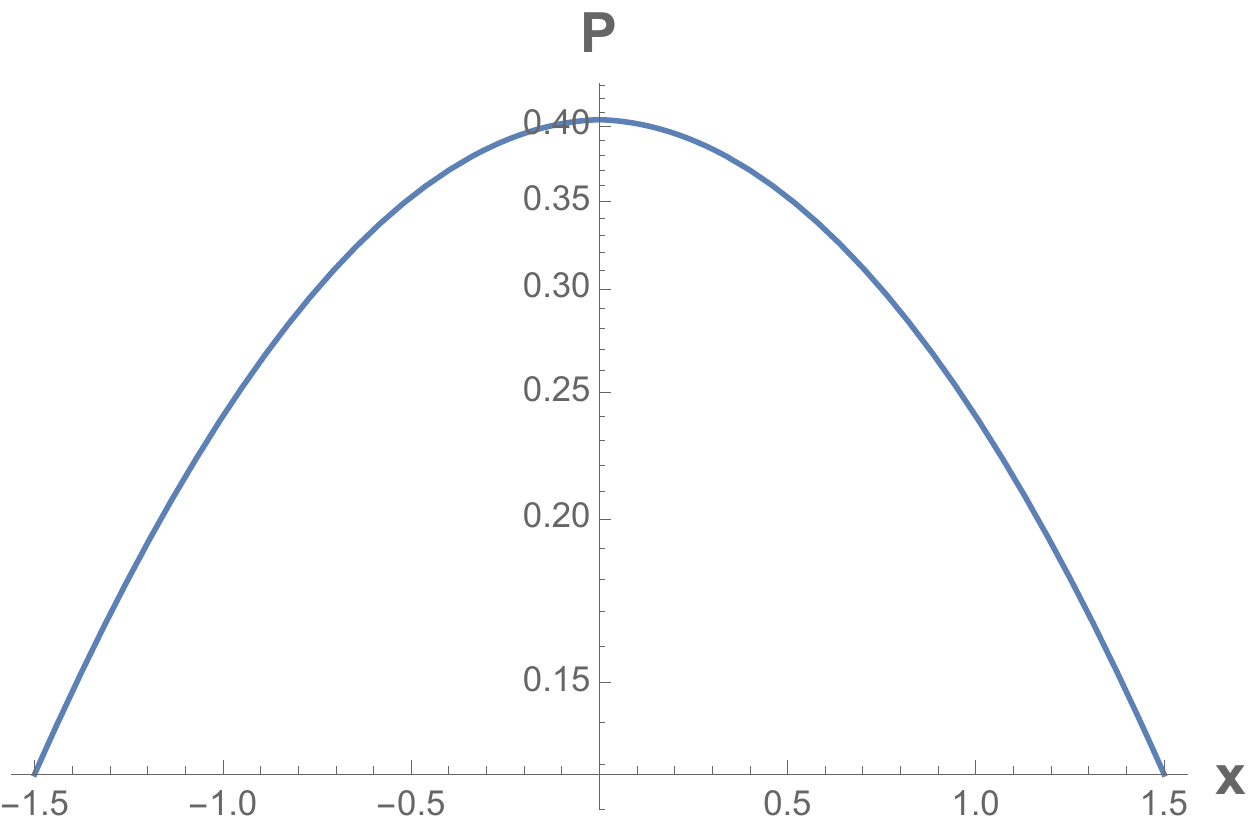}
\includegraphics[scale = 0.4]{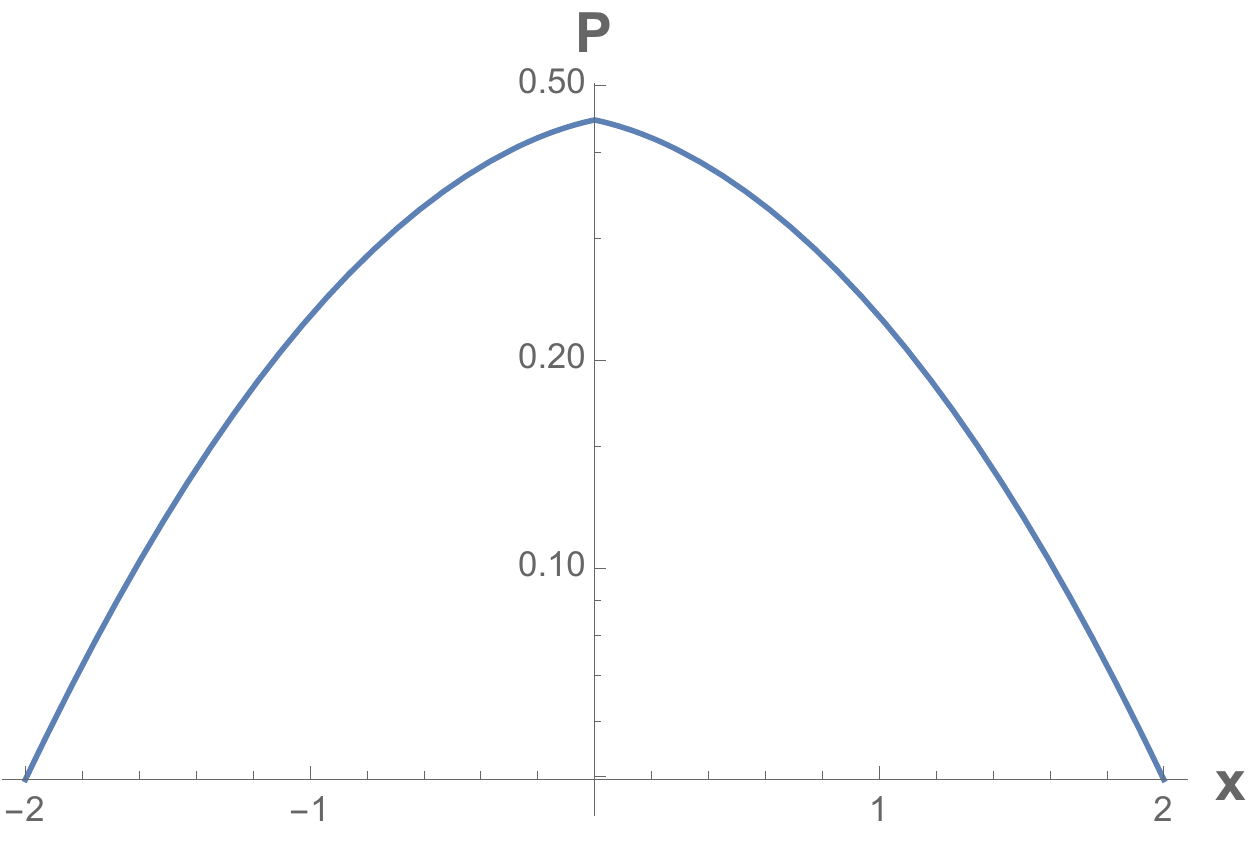}
\includegraphics[scale = 0.4]{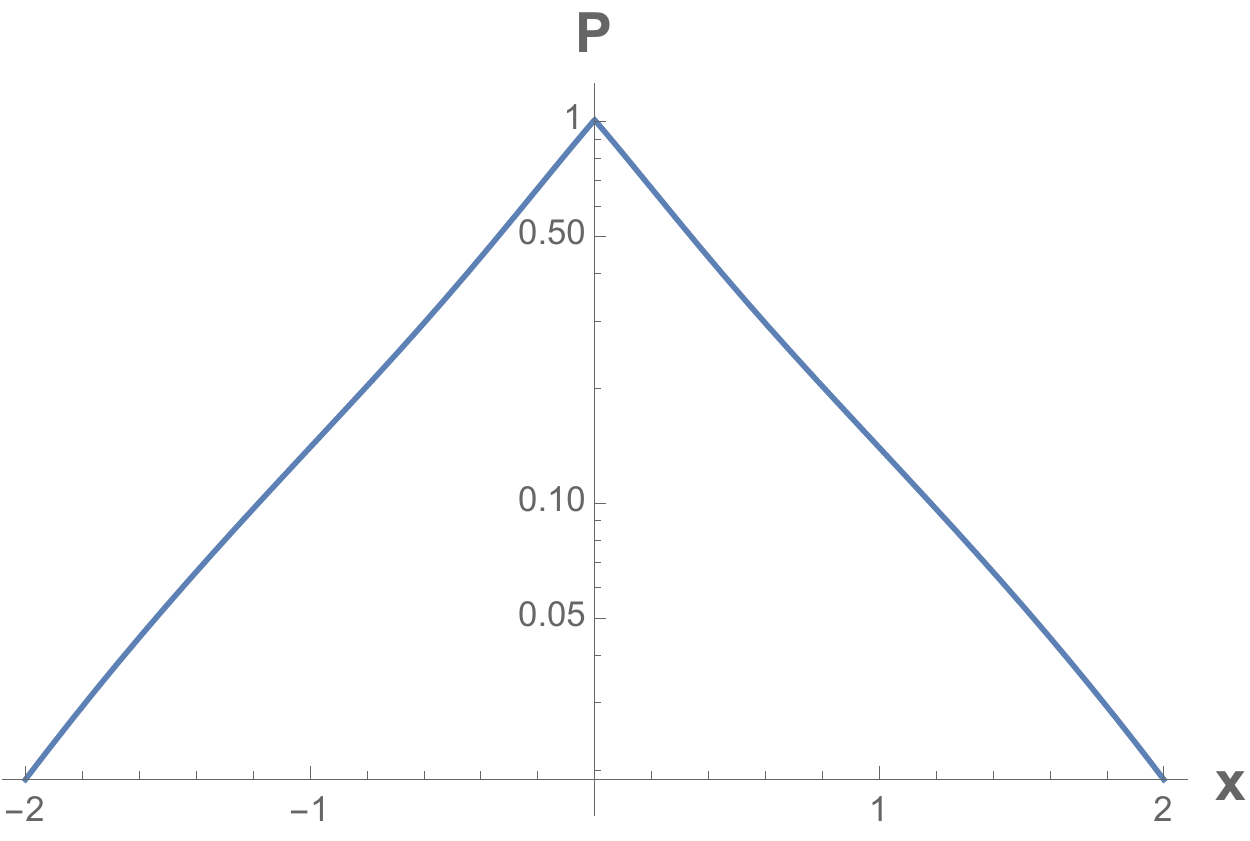}
\caption{ Probability  densities  of normalized dissipation rate $e={\cal E}/\overline{\cal E}$  vs   "Reynolds number" $y$.\\ 
Left  panel: $y=0.01$. Middle: $y=0.1$. Right: : $y=1.$. In all  cases $\beta=1.4$  as estimated in [12]-[13].  }
\label{fig1}
\end{figure}
 


\noindent {\bf This expression  is valid when Reynolds number is so small that the non-linearity  in (7)  can be neglected, but large enough   to allow for the  relatively weak 
boundary layer instability  leading to isolated (discrete)  plumes.}\\
In the interval  $y\ll X$,  $yX<X^{2}$ and the probability density $P(X)$  is close to the Gaussian with the first few low-order moments 
$e_{n}\propto \overline{X^{2n}}\approx (2n-1)!!$.  

\noindent It is interesting  that appearance of two "Reynolds numbers"     $Re$ and  $y\propto Re-Re^{tr}$ in (12) ,  reflects two different mechanisms  of turbulence production  experimentally observed by Tong et.al.  [15].  Indeed, when 
$y\rightarrow 0$, the length-scale-lacking-fluctuations  obey Gaussian statistics.   
One can also see, that,  as the  transitional ``Reynolds number"  $y$ grows,  so that $yX>X^{2}$,  the PDF  (12) varies to close- to -exponential.

\subsection{High Reynolds number limit. Mean-Field theory.}

\noindent In the limit  $Re\rightarrow\infty$ or $\nu\rightarrow 0$ and $\kappa\rightarrow 0$, far enough from thermal boundary layers
mean temperature $\Theta\approx const$ and the temperature gradient  $\frac{\partial \Theta}{\partial z}$ is very small.   Therefore,  according to (7)-(10) in dimensionless form:

$$S^{T}_{3}(r)\approx-\frac{4}{3}Nr; \hspace{2cm}
S_{3}(r)\approx \frac{2\alpha g}{r^{4}}\int_{0}^{r}y^{4}\overline{\delta_{r}v\delta_{r}T}dy$$

\noindent  A simple approximation, consistent with the mean-field theory developed below:

$$(\delta_{r}v)^{3}\approx 2\alpha g (\delta_{r}v)(\delta_{r}T)$$

\noindent gives : $\overline{(\delta_{r}T)^{2}}\propto N^{\frac{4}{5}}g^{-\frac{2}{5}}r^{\frac{2}{5}}$ and $S_{2}(r)\propto g^{\frac{4}{5}}N^{\frac{2}{5}}r^{\frac{6}{5}}$ leading to "Bolgiano-like"   energy spectra:

\begin{eqnarray}
E_{T}(k) \approx N^{\frac{4}{5}}g^{-\frac{2}{5}}k^{-\frac{7}{5}}; \hspace{1.5cm} E(k) \approx N^{\frac{2}{5}}g^{\frac{4}{5}}k^{-\frac{11}{5}}
\end{eqnarray}

\noindent  In this paper we consider the relatively simple case of $Pr=\nu/\kappa=1$ and  in the mean - field approximation,     scale-dependent ("dressed"),  transport coefficients,   derived in the one-loop renormalized Wild's diagrammatic expansion gives: 

$$\nu(k)=\sqrt{E(k)/k}\approx N^{\frac{1}{5}}g^{\frac{2}{5}}k^{-\frac{8}{5}};
 $$

\noindent 
In the "inertial range" of both energy and temperature, the  transport coefficients are related: $\nu(k)\approx \kappa(k)$. 
However in the dissipation range of turbulence $\nu_{T}(k) \rightarrow \kappa$ and at the dissipation scale $\nu(l_{d})\approx \nu=\kappa$

\noindent It follows from the above relations that in the inertial range the heat flux is:

$$J_{T}=\nu(k)\int_{0}^{k}k^{2}E_{T}(k)dk=N=const$$ 

\noindent This result means that in this range of scales,    heat  (temperature) is neither generated nor dissipated. While  production mechanism is dominated by convection rolls on the scale $r\approx H$, the heat is dissipated on a scale . 
$r\leq l_{d}^{T}$, calculated below. \\

\noindent  Also, we have  from (8)-(9) that at  the temperature dissipation scale  $l^{T}_{d}$:

$$\frac{4}{3}Nl^{T} _{d}\approx 2\kappa N^{\frac{4}{5}}g^{-\frac{2}{5}}(l^{T}_{d} )^{-\frac{3}{5}}$$

\noindent and

\begin{eqnarray}
l^{T}_{d}\approx \kappa^{\frac{5}{8}}N^{-\frac{1}{8}}g^{-\frac{1}{4}}
\end{eqnarray}

\noindent Repeating this calculation for the energy flux $J(k)$,  we,  using  the energy spectrum $E(k)$ from  (11), obtain :

$$J(k)=\nu(k)\int_{0}^{k}k^{2}E(k)dk =N^{\frac{3}{5}}g^{\frac{6}{5}}k^{-\frac{4}{5}}$$



\indent In the  dissipation range $r<l_{d}^{T}$,  the temperature fluctuations are very weak. At the same time, the velocity field,  stirred at the scale $r\approx l_{d}^{T}$,  is still strong. Indeed, the energy flux $J(k_{i})$ where $k_{i}\approx 1/ l_{d}^{T}$, is $O(1)$. 
\noindent Again, using turbulent diffusivity  $\nu(k)\approx 0.7\kappa(k)$ applied to the temperature spectrum $E_{T}(k)$, we derive: 
$$N=\overline{v_{3}T}[\frac{\partial \Theta}{\partial z}]= \kappa_{T}(\frac{1}{H})[\frac{\partial \Theta}{\partial z}]^{2}=\frac{J_{T}^{2}}{\kappa_{T}(1/H)}
 \approx J_{T}^{2}g^{\frac{1}{7}}/H^{\frac{6}{7}}$$
  \noindent where $1/k_{I}\approx H$ is the integral scale.

\noindent Since and $g=O(Ra)$ and  $N=O(Ra^{\frac{1}{7}} )$ we have from (12): 

\begin{eqnarray}
H/l_{d}^{T}\approx Nu \approx Ra^{\beta}
\end{eqnarray}

\noindent  where $\beta\approx 1/4+1/56\approx 0.27$ close to the result of Chicago experiment $\beta\approx 2/7\approx 0.28$ [18].
 This result, derived for a particular set up,  is in a generally  reasonable agreement with experimental data ranging in the interval $0.27\leq \beta  \leq 0.3$ with some logarithmic corrections.   At this point, the role  of  aspect ratio, strong intermittency, Prandtl and Rayleigh numbers  is hard to assess.\\

\noindent In the temperature dissipation range $r<l_{d}^{T}$ where  temperature fluctuations are overdamped, there exist an interval of  constant energy flux equal to: 

\begin{eqnarray}
J(l_{d}^{T})=N^{\frac{3}{5}}g^{\frac{6}{5}}(l_{d}^{T})^{\frac{4}{5}}\approx g^{\frac{17}{10}}\approx Ra^{\frac{17}{10}}
\end{eqnarray}

\noindent with  Kolmogorov's energy spectrum:

$$
E(k)\approx Ra^{\frac{17}{15}}k^{-\frac{5}{3}}
$$

\noindent The energy dissipation scale, calculated from the balance $\nu(k)=\sqrt{E(k)/k}=\nu_{0}\propto 1/Re\approx Ra^{-\frac{1}{2}}$  is thus:

\begin{eqnarray}
l_{d}\approx Ra^{-\frac{16}{15}}
\end{eqnarray}

From (8) -(11) one derives readily: $Re^{2}=O(v_{rms}^{2})\approx N^{\frac{2}{5}}g^{\frac{4}{5}}\propto Ra^{\frac{6}{7}}$  and $\frac{Ra}{Re^{2}}=O(Ra^{\frac{1}{7}})\rightarrow \infty$
used as an assumption leading to balances (8)-(9).   In the limit $Ra\rightarrow\infty$, the relations (11)-(12) give:

\begin{eqnarray}
\frac{l_{d}^{T}}{l_{d}}\approx Ra^{0.797}\rightarrow\infty
\end{eqnarray}


{\bf As we see, in the  temperature dissipation range  $r<l_{d}^{T}$,  where temperature fluctuations are negligibly weak,   there exist  an intermediate range  

$$l_{d}<<r<<l_{d}^{T}$$

\noindent of constant energy flux,  given by (20),   corresponding  to "Kolmogorov turbulence".   The most important consequence of this result is that in this range the turbulent kinetic energy  is produced by  the 
"large-scale" ($r\approx l_{d}^{T})$ forcing, which is the  temperature field  of eqs.(7)-(8).
\noindent The  dynamics of velocity derivatives at the scales  $l_{d}\ll l_{d}^{T}$, governed  by the equations  (7)-(8),  is reduced   to a familiar  problem of turbulence driven by the large-scale random force discussed above.  
  }

\section{Summary and discussion. Universality.}

\noindent  Numerical experiments of  Refs.[2]-[4]  on  infinite flows driven by two different large-scale  random forces  revealed an unexpected transition  from  a  low-Reynolds number Gaussian velocity field to  multi-scaling  
at the Reynolds number $R_{\lambda,2}\approx 8.91$. 
This, forcing-independent  result,  hints to a possible non-trivial universality  or universality classes of turbulent flows. To test this assumption,  in this paper, we study the process of emergence of anomalous scaling in  Rayleigh-Benard convection. It is shown that in this system,  the temperature field acts as a {\bf large-scale} stirring force for the velocity field and it  has been shown analytically (13) that in the vicinity of transition point $R_{\lambda}\approx 9.0$, this force generates a  Gaussian random flow, similar to that in homogeneous and isotropic turbulence of Refs.[2]-[4].  Thus, the anomalous scaling emerges at $R^{tr}_{\lambda,2}\approx 8.91$ calculated in Ref.[6]-[10], in a  proximity of  a
 Gaussian field point.   While   in the high Reynolds number 
limit  $R_{\lambda} \gg  9.0$,  all three flows are characterized by the same multi-fractal exponents, they may be very different in the
weak-coupling or linear  limit  $R_{\lambda} <   9.0$ .  
 Recently, Das and Girimaji (Ref.[4]), investigating a  flow of  Refs.[2], discovered a sharp transformation  of some geometric flow characteristics  at $R_{\lambda}\approx 9.0$,  providing  a dynamic  interpretation  of   the observed transition.\\
 
 \noindent To study the dependence of a flow on the renormalized ("dressed") Reynolds number, one can start  at $R_{\lambda}\gg 9.0$
 and gradually  decreasing the Reynolds number toward $R_{\lambda}=R^{tr}_{\lambda}=9.0$,  follow emergence of a Gaussian 
 PDF of a temperature field. This  gaussian point of turbulence may explain successful one-loop application of Wyld's  renormalized perturbation expansion for derivation of turbulence models  widely used for simulation of complex engineering flows.  \\
 
 \noindent  All above calculations addressed a random flow generated in the central part of the cell, far enough from the wall-boundary layers.
 According to our assumption, due to efficient  mixing,  there,   
 $\frac{\partial \Theta}{\partial z}$ is small and,  therefore,  the heat flux  contribution to the balance relation  (9) is negligibly small.
 Corresponding  contribution to the balance relation (10) is large,  leading to  very different energy and heat spectra.  This results in heat and energy fluxes operating at widely different length-scales intervals with  the energy dissipation scale $l_{d}\ll l^{T}_{d}$. Thus, the temperature field acts as a large-scale stirring force for the velocity field. This brings the small-scale  velocity flow  in RBC  into a class 
 of homogeneous and isotropic turbulence considered in Ref.[2]-[4].\\
  
 \noindent Now, we would like to address Gaussianity of the low-Reynolds number velocity field in the vicinity  transition point $R_{\lambda}\approx 9.0$. The representation (12),  an {\bf exact}  consequence of equations  of motion (7)-(8),  is not a closed 
 equation for the PDF $P(X)$.  However, in the vicinity of transition point, when $X$ is small enough,  the low-order Taylor expansion gives (13) independently of the geometric  details. Increasing $X$ or a transition parameter $y$ leads, first,  to  close-to-exponential and even stretched exponential  tails.    These features of  RBC flow have been predicted in Ref.[12] and experimentally observed in Ref.[15] ,[17].
 
  \noindent The mean-field theory, developed above, gave $Nu\approx \frac{H}{l_{d}^{T}}\propto Ra^{\frac{15}{56}}$,   close to the result reported in Chicago experiment Ref.[17]. Modern data seem to be  closer to $Nu\propto Ra^{\beta}$ with $\beta\approx 0.29-0.31$,  with  some  dependence of exponent $\beta$ on the aspect ratio ${\Gamma}$ and Prandtl number  (Pr)  (see Ref.[14]).  Recently, in a high  quality numerical experiments on RBC in a very thin cell ( $\Gamma=d/H=0. 1$),   K. Iyer, et.al. ( Ref.[16]),  reported $\beta=0.331$,  very close to  a classic exponent derived  by Malcus in 1954 (Ref.18]). With increase of  the aspect ratio $\Gamma$,  the observed exponent was in the range $\beta\approx 0.28-0.30$ Ref.[14].  The present   mean-field theory  was developed for  the  flow in a cell of $\Gamma\rightarrow\infty$  which may,  after all,  be close to the asymptotic limit $\beta\approx 2/7$.   It still remains to be seen.

 \noindent It has to be stressed that in this work no  effects of  intermittency have been accounted for as well as  the role of Prandtl number,  aspect ratio, instabilities of side walls boundary layers which, in principle,  can explain the small  difference between our mean-field theory and experimental data. Another possibility is much more exciting: according to  the  developed  above theory,  the transition point  $R_{\lambda}\approx 8.91$ is Gaussian,  which may explain validity of the mean-field approximation and $\beta\approx 15/56$ in a flow between infinite plates and $Pr=1$,  which is beyond present-day experimental setups. Today, this suggestion can be considered as  a  mere speculation.

\section*{Acknowledgements.} 

\noindent  At the early stages leading to this paper, I benefitted a lot from collaborations wth S.A.Orszag and Ya.G. Sinai.  I  am profoundly grateful to  D.Donzis, K.R.Sreenivasan,  J.Schumacher, R.Das and S.Girimaji, L.M.Smith for  discussions of various aspects of the problem.  Also,  the input of Drs. Chen and Staroselsky   
 of EXA Corporation for 
 sharing  a lot of data on ``turbulent''  Reynolds numbers in various applications.


\begin{thebibliography}{1}


\bibitem{Yaglom}
A.M. Yaglom,
{\em  Hydrodynamic Instability and Transition to Turbulence, 
Springer, ,2012; \ 
L.E.Lorenz,  ``Deterministic nonperiodic flow'', J.Atmos.Sci. {\bf 20}, 130-141 (1963 ).}



\bibitem{Yakhot1986}
J. Schumacher, K. R. Sreenivasan, and V. Yakhot,
New J. Phys. {\bf 9}, 89 (2007).


\bibitem{Yakhot2017}
V. Yakhot and D. A. Donzis,
Phys. Rev. Lett. {\bf 119}, 044501, (2017) 


\bibitem{Yakhot2018}
V. Yakhot and D. A. Donzis,
PhysicaD. {\bf xx}, 044501 (2018); \
 R.Das and S.Girimaji, 
``On the Reynolds number dependence of velocity structure 
and dynamics,"
J.Fluid.Mech.  2018 (in press)).

\bibitem{Landau1980}
L. D. Landau and E. M. Lifshitz,
{\em Course of Theoretical Physics, Statistical Physics, Volume 5},
Butterworth-Heinemann, Oxford, 1980.


\bibitem{launder}
B.E.\ Launder and D.B.\ Spalding, 
{``Mathematical Models of Turbulence'',  Academic
Press, New York (1972);  In turbulence modeling literature  ${\cal K} -{\cal E}$ model is used with experimentally determined coefficient $C_{\mu}=0.09$ instead of the derived $C_{\mu}=0.0845$.}


 \bibitem{yakhot1992} 
 V.\ Yakhot and L.\ Smith, 
{ ``The renormalization group, the $\epsilon$-expansion and derivation of
 turbulence models'',}
 J.\ Sci.\ Comp.\ {\bf 7}, 35 (1992).
 \bibitem{yakhot212}
V.~Yakhot,  
``Reynolds number of transition and self-organized criticality of strong turbulence'',
Phys.\ Rev.\ E,{\bf 90}, 043019 (2014). 

\bibitem{yakhot213}
 V.~Yakhot, S.A.\ Orszag, T.\ Gatski, S.\ Thangam and C.\
Speciale, 
{``Development of turbulence models for shear flows by a double expansion technique'',}
Phys.\ Fluids A{\bf 4}, 1510  (1992);  

\bibitem{Yakhot1986}
V.Borue, S.A. .Orszag, \
{``Local energy flux and subgrid-scale statistics in three-dimensional turbulence.'' }
 J.Fluid.Mech, {\bf 366},1,(1998)
 



 

\bibitem{Yakhot1989}
Ya.G. Sinai, and V. Yakhot,  
Phys. Rev. Lett. {\bf 63}, 1965 (1989).
V. Yakhot, Phys. Rev. Lett. {\bf 63}, 1965 (1989). 

\bibitem{Yakhot1990}
V. Yakhot, S. A. Orszag, S. Balachandar, E. Jackson, Z.-S. She, and L. Sirovich,
J. Sci. Comput. {\bf 5} (3), 199 (1990).


\bibitem{krishnamurti}
R. Krishnamurti,   
J . Fluid Mech. 33 457-631970a;
J. Fluid Mech. 42 295-307 1970b; 

\bibitem{busse1978}
Busse, F. H.,  
 {Non-linear properties of thermal convection}
 1978 Rep. Prog. Phys. 41  1929 p.1943;


\bibitem{Tong2009}
X.He and P.Tong,
{``Measurements of thermal dissipation field in Benard convection''}
Phys.Rev.E,{\bf 79},026306 (2009)


\bibitem{Schumacher}
 Kartik P.Iyer, J.D.Scheel, J. Schumacher  and K.R.Sreenivasan, 
{``Classical 1/3 scaling of convection holds up to $Ra=10^{15}$''}
PNAS, (2020))

\bibitem{Castaing1989}
B. Castaing, G. Gunaratne, F. Heslot, L. P. Kadanoff, A. Libchaber, S. Thomae, X.-Z. Wu, S. Zaleski, G. Zanetti,
J. Fluid Mech. {\bf 204}, 1 (1989).
 
 


 
\bibitem{Malkus1954}
W. V. R. Malkus, 
Proc. R. Soc. Lond. A {\bf 225}, 185 (1954).

 
\end{thebibliography}
\end{document}